# Crystal fields and Kondo effect: new results for the magnetic susceptibility

H.-U. Desgranges*
*Scientifically unaffiliated, Albert-Kusel-Str. 25, Celle, Germany*

The thermodynamic Bethe ansatz equations for the Coqblin-Schrieffer model have been solved for the first time to obtain the magnetic susceptibility in the presence of crystal fields for non-zero temperatures. For the case of N = 4 effective ionic states an analytic expression for the limiting values of the pseudo-energies has been found facilitating the numerical solution for various crystal and magnetic field configurations. The single-impurity model applies to a wide range of dense Kondo systems and has been used before to explain apparent non-Fermi-liquid behavior. The flattening off of the susceptibility curves at a substantially higher temperature than the specific heat is shown to be a general feature of the Coqblin-Schrieffer thermodynamics.

## 1. Introduction

There is an increasing amount of experimental data on heavy fermion compounds, showing a great variety in their behavior, which is still far from being understood. One particular aspect that deserves further clarification is the relative effects of the crystal field and the Kondo scattering terms in the presence of magnetic fields.

The single-impurity Kondo and Anderson models and their generalizations to include orbital degeneracy serve as testing grounds for methods that are aimed at the corresponding lattice problems. The Bethe ansatz solutions to the impurity models in turn provide benchmarks [1-5] for ground state and finite-temperature thermodynamic quantities. The numerical solution to the thermodynamic Bethe ansatz (TBA) equations is able to render the full crossover between local-moment behavior at high temperatures and Fermi-liquid behavior at low temperatures with relatively little numerical effort.

On the experimental side there is on-going interest in the interplay of crystal fields and the Kondo effect [6-10]. The experimentalists' analysis of the measured thermodynamic quantities (for temperatures larger than a magnetic transition as the case may be) has been performed essentially in two steps: The low temperature behavior is fitted to the exact result for the spin-1/2 Kondo model [11]. Here the analytically available result for the resonant level model may be taken as an approximation. The higher temperature region is fitted by the Schottky curve for the non-interacting crystal field Hamiltonian. Recently an attempt has been made [12] to combine the resonant level and crystal field approaches to cover the whole applicable temperature range.

On the theoretical side, the generalization of the single ion Kondo model to a N-fold degenerate ionic configuration, the SU(N) Coqblin-Schrieffer [13] model was solved by Bethe ansatz more than 30 years ago [14]. An overview over the results for dilute mixed-valent and heavy-fermion systems derived from the TBA equations is given in ref. [15]. An alternative approach to the solution of the Coqblin-Schrieffer model is presented in ref. [16]. There, formulae are given for the calculation of weak field and low temperature expansions of the free energy of the model. A broad basis for comparison with experiments on the specific heat in zero magnetic field over the whole temperature range has been provided recently by the numerical solution of the TBA equations for the N = 6 model (Cerium $3^+$ ions) with general crystal field configurations [17]. A new high field / low temperature expansion was developed there to calculate the limiting values of the unknown functions as a basis for the numerical solution.

An interesting aspect of the interplay between degeneracy and crystal fields has been raised by Anders and

*E-mail address: H-Ulrich.Desgranges@nexgo.de

Pruschke [18] who studied the problem by the numerical renormalization group method for the case of N = 4 and calculated specific heat and magnetic susceptibility to analyze experimental data on $Ce_{1-x}La_xNi_9Ge_4$. These authors argue that the apparent non-Fermi-liquid behavior can be explained by an extended crossover regime caused by the crystal field that leads to the flattening off of the susceptibility curves at a substantially higher temperature than the specific heat. Moreover, they claim that this holds only for a narrow range of crystal field splittings.

In contrast to that assertion I show that this behavior is a general feature of the Coqblin-Schrieffer model thermodynamics for arbitrary crystal field splittings.

Following Anders and Pruschke I examine the case of N = 4 that can hold as an approximation to the physical N = 6 ($Ce\ 3^+$ ions) or N = 8 ($Yb\ 3^+$ ions) case when a low lying quartet or two low lying doublets are separated from the higher multiplets so much that the upper multiplets can be neglected in the low temperature thermodynamics.

I consider three cases of crystal and magnetic field environments that are determined by crystal symmetry and the orientation of the magnetic field. When comparing with experiments the value of the Landè factor g has to be taken as that of the full Hund's rule ionic configuration without crystal fields:

(a) An (effective) spin J = 3/2 ion in a uniaxial crystal field [19]. Without magnetic field the quartet is split into two doublets separated by an energy difference Δ. The energy levels with magnetic field applied along the z-axis may be labeled from 1 to 4 and are given by:

$$E_1 = -\tfrac{1}{2}\Delta - \tfrac{1}{2}g\mu_B H, \quad E_2 = -\tfrac{1}{2}\Delta + \tfrac{1}{2}g\mu_B H, \quad E_3 = +\tfrac{1}{2}\Delta - \tfrac{3}{2}g\mu_B H, \quad E_4 = +\tfrac{1}{2}\Delta + \tfrac{3}{2}g\mu_B H. \tag{1}$$

(b) A $\Gamma_8$ quartet (i.e. cubic environment) with tetragonal distortion [20] with the magnetic field applied along the fourfold axis. The energy levels are given by:

$$E_1 = -\tfrac{1}{2}\Delta - \tfrac{1}{2}g\mu_B H, \quad E_2 = -\tfrac{1}{2}\Delta + \tfrac{1}{2}g\mu_B H, \quad E_3 = +\tfrac{1}{2}\Delta - \tfrac{11}{6}g\mu_B H, \quad E_4 = +\tfrac{1}{2}\Delta + \tfrac{11}{6}g\mu_B H. \tag{2}$$

(c) A configuration considered in ref. [18] and [21] where the relative g-factor $g_{rel}$ between the two doublets is determined to fit the experiments for $Ce_{0.5}La_{0.5}Ni_9Ge_4$ and is equal to √2. The energy levels are given by:

$$E_1 = -\tfrac{1}{2}\Delta - \tfrac{1}{2}g\mu_B H, \quad E_2 = -\tfrac{1}{2}\Delta + \tfrac{1}{2}g\mu_B H, \quad E_3 = +\tfrac{1}{2}\Delta - \tfrac{1}{2}g_{rel}g\mu_B H, \quad E_4 = +\tfrac{1}{2}\Delta + \tfrac{1}{2}g_{rel}g\mu_B H. \tag{3}$$

The rest of this publication is organized as follows: In Section 2 the model is introduced and the TBA equations are formulated. The results for the limiting values needed to solve these non-linear integral equations are given in Sec. 3. An overview over the numerical results on the magnetic susceptibility for high and low temperatures is given in Sec. 4 exemplarily for case (a). From these results the zero temperature values of the magnetic susceptibilities for the three cases considered are extracted and shown as functions of the splitting Δ. In Sec. 5 results for representative values of the crystal field splittings are provided for all three cases to compare with experimental data. The apparent non-Fermi-liquid behavior is shown to be a general feature of the Coqblin-Schrieffer thermodynamics. The behavior of the limiting values of the unknown functions of the TBA equations at large values of the equation index n is given in the appendix.

## 2. Model and TBA equations

The Coqblin-Schrieffer Hamiltonian can be written in terms of the N ionic crystal field states $|r\rangle$ with energy levels $E_r$ and the usual notation for conduction electron operators $C^{\dagger}_{k,r}$. The exchange interaction is simply a permutation operator acting on the quantum labels of the particles.

$$H = \sum_{k,r} k\, C^{\dagger}_{k,r} C_{k,r} + J \sum_{k,r;\,k',r'} |r\rangle\langle r'|\, C^{\dagger}_{k',r'} C_{k,r} + \sum_r E_r\, |r\rangle\langle r| \tag{4}$$

For integrability of the model a linear dispersion of the conduction electron energy is assumed as well as the smallness of the exchange coupling *J* independent of *r*. The Bethe ansatz solution introduces an ad hoc cut-off D that enters the Kondo temperature $T_K \sim D \exp(-1/N|J|)$ in a non-universal way [11]. In the scaling limit $J \to 0$, $D \to \infty$, $T_K$ is kept fixed and is the only scale of the model whose value may be fitted to the experiments.

In the language of the Anderson model the conditions on the crystal fields reduce to the requirement that the size of the splittings $E_{r+1} - E_r$ be negligible compared to both the bare f-level position and the conduction-electron bandwidth. The splittings may then be large or small compared with the Kondo temperature.

The thermodynamic properties of the model are calculated from certain pseudo-energy functions $\varepsilon_n^{(r)}(\lambda)$, $n = 1, 2, ..., \infty$, $1 \le r \le N-1$ that are determined by the TBA equations [14] (with $\varepsilon_0^{(r)} = -\infty$):

$$-\ln\{\, 1 + \exp[-\varepsilon_n^{(r)}(\lambda)/\mathrm{T}]\,\} = -\sin(r\,\pi\,/\,N)\,\exp[\lambda]\delta_{n,1}$$
$$+\sum_{q=1}^{N-1} S_q^r * (\ln\{1+\exp[\varepsilon_{n+1}^{(q)}(\lambda)/\mathrm{T}]\} + \ln\{1+\exp[\varepsilon_{n-1}^{(q)}(\lambda)/\mathrm{T}]\} - s^{-1} * \ln\{1+\exp[\varepsilon_n^{(q)}(\lambda)/\mathrm{T}]\}). \tag{5}$$

Here $s * f(\lambda)$ denotes the convolution $s * f(\lambda) = \int_{-\infty}^{\infty} s(\lambda-\lambda')\, f(\lambda')\, d\lambda'$, and the kernels $S_q^r$ are given by their Fourier transforms:

$$\mathrm{S}_q^r(\omega) = \frac{\sinh(\min(q,r)\pi\omega/N)\ \sinh((\mathrm{N}\text{-}\max(q,r))\pi\omega/N)}{\sinh(\pi\omega)\,\sinh(\pi\omega/N)} \quad \text{and} \quad \mathrm{s}^{-1}(\omega) = 2\cosh(\pi\omega/2)\,.$$

The free energy at temperature T is given by the following expression:

$$\mathrm{F} = -\mathrm{T}\sum_{r=1}^{N-1}\int_{-\infty}^{\infty} \frac{\sin(r\pi\,/\,N)\,\ln\{\,1+\exp[\varepsilon_1^{(r)}(\lambda\,)/\mathrm{T}]\,\}\,\mathrm{d}\lambda}{\{\cosh[\lambda - \ln(\mathrm{T_K}\ /\ \mathrm{T})] - \cos(r\,\pi\,/\,N)\}\,2\pi} \tag{6}$$

The thermodynamic properties depend only on the ratio $T/T_K$ and on the external fields scaled by $T_K$. The definition of the Kondo temperature $T_K \equiv T_K(N)$ used here connects it with the linear specific heat coefficient in the absence of all fields $\gamma_0 = C/T$ for $T \to 0$ through $T_K(N) = (N-1)\,\pi/(3\gamma_0)$.

By introducing $g_n^{(r)}(\lambda) \equiv \ln\{1+\exp[\varepsilon_n^{(r)}(\lambda)/\mathrm{T}]\}$ the equations (5) can be written for $n \ge 2$ in the following form (with $g_n^{(0)} = g_n^{(N)} \equiv \infty$):

$$g_n^{(r)} = s * g_{n+1}^{(r)}(\lambda) + s * g_{n-1}^{(r)}(\lambda) + \ln\{1 - \exp[-g_n^{(r)}]\}$$
$$+ s * \ln\{1 - \exp[-g_n^{(r+1)}(\lambda)]\} + s * \ln\{1 - \exp[-g_n^{(r-1)}(\lambda)]\}. \tag{7}$$

The numerical solution of eqs. (7) together with eqs. (5) for n = 1 is facilitated by the knowledge of the limiting values of the functions $g_n^{(r)}(\lambda)$ for $\lambda \to \pm\infty$.

With the notation: $\bar{g}_n^{(r)} \equiv g_n^{(r)}(-\infty) = g_{n+1}^{(r)}(+\infty)$, $b_n^{(r)} \equiv -\ln\{\,1 - \exp[-\bar{g}_n^{(r)}]\}$ the integral equations in this limit reduce to algebraic recurrence relations:

$$\bar{g}_n^{(r)} - \tfrac{1}{2}\{\bar{g}_{n+1}^{(r)} + \bar{g}_{n-1}^{(r)}\} = b_n^{(r)} - \tfrac{1}{2}\{b_n^{(r+1)} + b_n^{(r-1)}\},\ \bar{g}_0^{(r)} = 0,\ \ b_n^{(0)} = b_n^{(N)} = 0\,, \tag{8}$$

$$\lim_{n\to\infty} \bar{g}_{n+1}^{(r)} - \bar{g}_n^{(r)} = \mathrm{A_r}/\mathrm{T}\,. \tag{9}$$

The generalized fields $A_r$ ($A_r \ge 0$) are related to the energy levels $E_r$ of the ionic ground state in the particular crystal and magnetic field: $A_r = E_{r+1} - E_r$, $1 \le r \le N-1$ with:

$$E_1 = -\tfrac{1}{2}\Delta - \tfrac{1}{2}g\mu_B H,\quad E_2 = -\tfrac{1}{2}\Delta + \tfrac{1}{2}g\mu_B H,\quad E_3 = +\tfrac{1}{2}\Delta - \tfrac{3}{2}\hat{g}g\mu_B H,\quad E_4 = +\tfrac{1}{2}\Delta + \tfrac{3}{2}\hat{g}g\mu_B H, \tag{10}$$

leading to:

$$A_1 = g\mu_B H, \qquad A_2 = \Delta - \frac{3\hat{g}+1}{2} g\mu_B H, \qquad A_3 = 3\hat{g}g\mu_B H. \tag{11}$$

The relative g-factors $\hat{g}$ in this formulation are equal to 1 in case (a), 11/9 in case (b) and √2/3 in case (c), respectively. Furthermore, since we are interested in the magnetic susceptibility at zero magnetic field I consider only the case that the Zeeman splitting $g\mu_B H$ is smaller than the crystal field splitting Δ such that $A_2 > 0$ is assured. As a consequence the limits H → 0 and Δ → 0 may not be interchanged.

In the absence of the exchange coupling i.e. for non-interacting spins in a crystal field the magnetic susceptibility can be calculated from the partition function and leads to the Curie law $\chi(T) = C/T$ at high T with the Curie constant C given by $a \cdot \tfrac{5}{4}(g\mu_B)^2$ where the factor $a \equiv (9\hat{g}^2 + 1)/10$ characterizes the influence of the crystal field symmetry on the Zeeman splitting.

In order to calculate the magnetic susceptibility in zero magnetic field $\chi(T) = -\,\partial^2 F(H,T) / \partial^2 H^2 \big|_{H=0}$ it is expedient to calculate the ancillary functions [22, 23] $E_n^{(r)}(\lambda) = \partial^2 g_n^{(r)}(\lambda) / \partial^2 (H/T)^2 \big|_{H=0}$. Building on the functions $g_n^{(r)}(\lambda)$ at H = 0 these are determined by the following equations for n≥2:

$$\begin{aligned} E_n^{(r)}(\lambda) &= \{1 - \exp[-g_n^{(r)}(\lambda, H=0)]\} \cdot s * \{E_{n+1}^{(r)}(\lambda) + E_{n-1}^{(r)}(\lambda)\} \\ &\quad + \sum_{\pm} s * \{E_n^{(r\pm1)}(\lambda) \cdot \exp[-g_n^{(r\pm1)}(\lambda, H=0)] / (1 - \exp[-g_n^{(r\pm1)}(\lambda, H=0)])\}. \end{aligned} \tag{12}$$

For n=1 one obtains:

$$E_1^{(r)}(\lambda) = \{1 - \exp[-g_1^{(r)}(\lambda, H=0)]\} \cdot \{E_1^{(r)}(\lambda) + \sum_{q=1}^{N-1} S_q^r * (E_2^{(q)}(\lambda) - s^{-1} * E_1^{(q)}(\lambda))\} \tag{13}$$

The zero field magnetic susceptibility can thus be calculated as:

$$\chi(T) = T^{-1} \sum_{r=1}^{N-1} \int_{-\infty}^{\infty} \frac{E_1^{(r)}(\lambda)\,\sin(r\pi/N)\,d\lambda/(2\pi)}{\cosh[\lambda - \ln(T_K/T)] - \cos(r\pi/N)} \tag{14}$$

## 3. New results for the limiting values

The limiting values $\bar{g}_n^{(r)}$ at H = 0 have been determined by a new method [17] that yields analytic expressions for nΔ/T >> 1 (see Appendix) that allow the numerical calculation for all n. The Δ/T-dependence of the results has been analyzed by means of a regression analysis in powers of a new variable x ≡ tanh(Δ/2T). Together with the required limiting behavior for small and large crystal fields this has led to the conjecture:

$$\begin{aligned} \bar{g}_1^{(1)} &= \bar{g}_1^{(3)} = \ln(8/3) - \ln(1 - x^2/3) \\ \bar{g}_1^{(2)} &= \ln(\frac{9}{4}) + \ln\{1 + \frac{1}{3}\frac{x^2}{1-x^2}(1 + \frac{1}{3}x^2)\} = \frac{\Delta}{T} + \ln(\frac{9}{4}) + 2\ln\{\frac{1 - x^2/3}{1+x}\}. \end{aligned} \tag{15}$$

Both functions agree with the numerically obtained values to an accuracy of $10^{-8}$. It should be noted that this is the first time an analytic expression is given for limiting values of the TBA equations in the presence of unequal generalized fields $A_r$. However, it has not proved possible to find an analytic expression for $\overline{g}_n^{(r)}$ for general values of n.

An analogous result for the limiting values to the ancillary functions $\overline{E}_n^{(r)} \equiv E_n^{(r)}(-\infty) = E_{n+1}^{(r)}(+\infty)$ has been obtained for n = 1:

$$\overline{E}_1^{(1)} = \frac{5}{2}\{-a-(a-\tfrac{1}{5})x+\frac{4a}{3-x^2}\},$$
$$\overline{E}_1^{(2)} = \frac{5}{2}\{3a \qquad -\frac{8a}{3-x^2}\}, \tag{16}$$
$$\overline{E}_1^{(1)} = \frac{5}{2}\{-a+(a-\tfrac{1}{5})x+\frac{4a}{3-x^2}\}.$$

Due to the properties of the integral kernel in eq. (14) that expression can be evaluated exactly in the limit $T \to \infty$ provided the values $E_1^{(r)}(-\infty)$ are known. In this way eq. (16) is validated since the Curie constant comes out correctly as that for non-interacting spins with Zeeman splittings given by eq. (10):

$$\lim_{T\to\infty} T\cdot\chi = \sum_{r=1}^{N-1}\frac{4-r}{4}\overline{E}_1^{(r)} = (g\mu_B)^2\ \frac{5}{4}\ a\ . \tag{17}$$

## 4. Results for the magnetic susceptibility

With the results for the limiting values for $g_n^{(r)}(\lambda)$ and $E_n^{(r)}(\lambda)$ the numerical solution of the thermodynamic Bethe ansatz equations has been achieved for the three cases of crystal field and magnetic field environments considered and for a number of crystal field splittings. The relative accuracy is generally better than 1% with slightly larger deviations at zero temperature possible. The crystal field splitting Δ as well as the temperature T are scaled by the Kondo temperature in the absence of crystal fields $T_K(N=4)$.

Results for the effective moment $\chi{\cdot}T$ for case (a): spin J = 3/2 in a uniaxial crystal field are shown in Fig. 1. The curves for cases (b) and (c) look similar when the high temperature limits are scaled by eq. (17). As is well known for the case without crystal fields [11] the free spin values are reached slowly for $T \to \infty$ with large logarithmic corrections. However, for comparison with experiments containing for example Cerium $3^+$ ions it has to be borne in mind that the present model is valid only at low and intermediate temperatures where the upper crystal field doublet may be neglected.

For low temperatures and large crystal field splittings the results can be fitted by the curve for the SU(2) model with the effective Kondo temperature given by the scaling relation [24]:

$$T_K\left(SU\left(2\right)\right) = \frac{\pi}{3}\frac{T_K(N=4)}{\Delta}\frac{T_K(N=4)}{3.496} \tag{18}$$

For completeness the corresponding curves for the magnetic susceptibility are also shown.

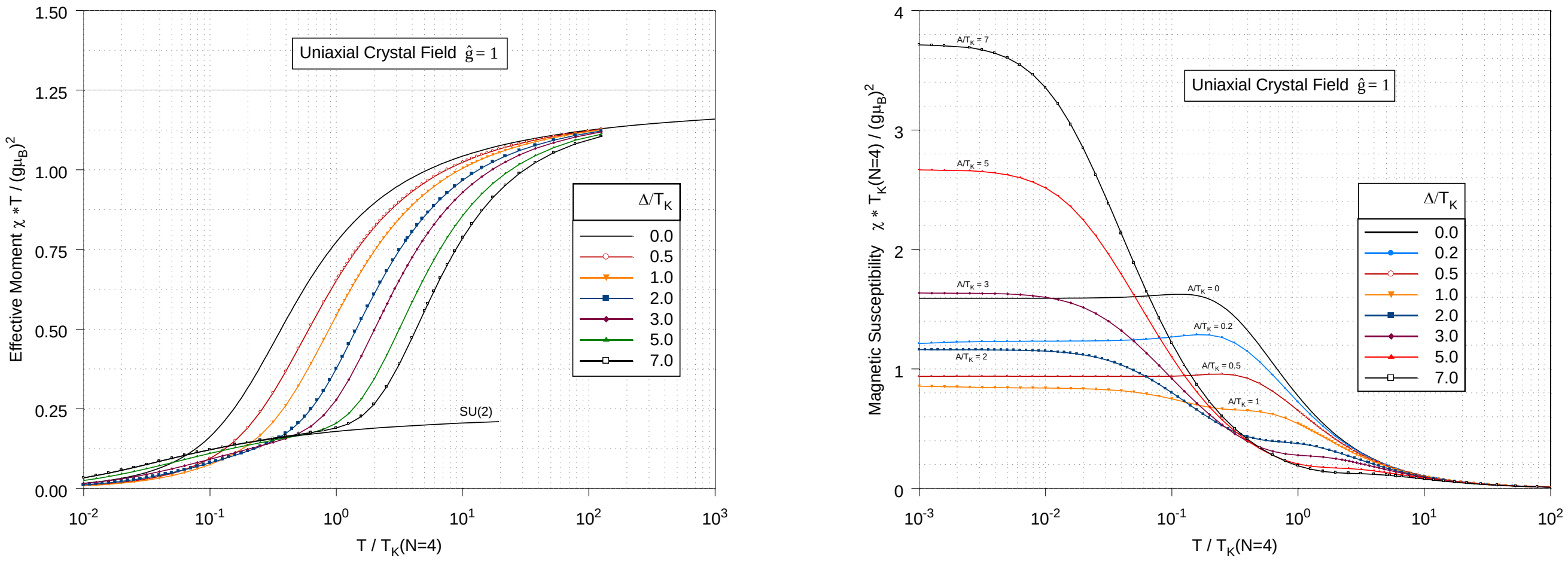

Fig. 1. (Color online) The effective moment $\chi \cdot T$ and the magnetic susceptibility $\chi$ as a function of T.

In Fig. 2 the dependence of the magnetic susceptibility at zero temperature on the crystal field splitting for the three crystal- and magnetic field configurations considered is shown.

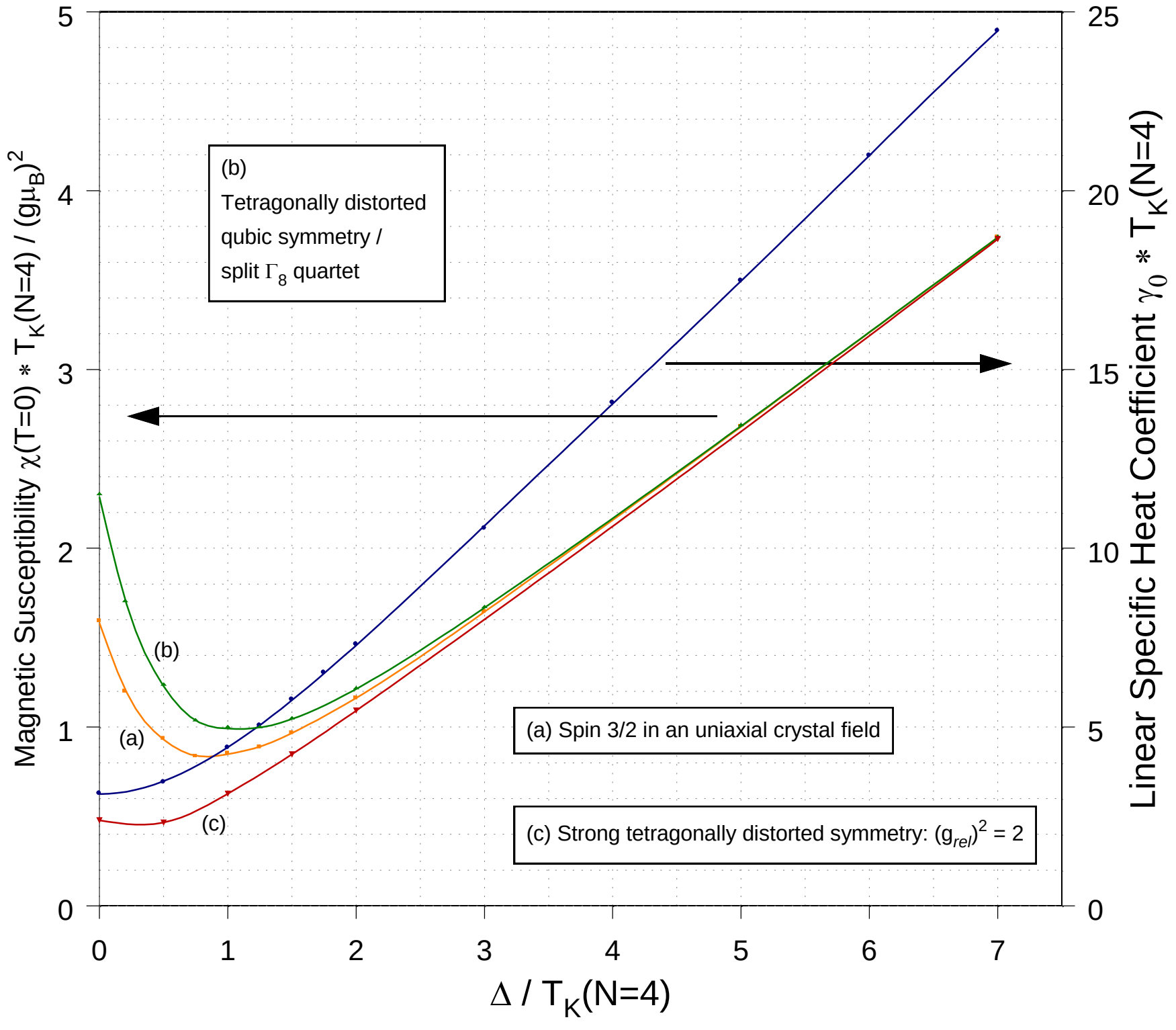

Fig. 2. (Color online) Zero temperature susceptibility $\chi(T=0)$ and linear specific heat coefficient $\gamma_0$ as functions of splitting $\Delta$.

Also shown is the linear specific heat coefficient $\gamma_0$ calculated in ref. [24]. As opposed to $\gamma_0$ the zero temperature magnetic susceptibility shows a minimum for non-zero values of the crystal field splitting. A similar behavior was found in ref. [25] where the zero temperature susceptibility for the N = 6 model with axial crystal fields was calculated. The position of this minimum is shifted towards larger values of $\Delta/T_K(N{=}4)$ for increasing values of $\hat{g}$ or *a* respectively. From eqs. 17-19 of ref. [16] one can calculate the zero temperature magnetic susceptibility to lowest order in the crystal field splitting $\Delta$:

$$\frac{T_K(N{=}4)}{(g\mu_B)^2}\chi(T{=}0) = \frac{5a}{\pi} - \frac{\Gamma^2(1/4)}{16\pi^3\sqrt{2\pi}}(18\hat{g}^2-1)\frac{\Delta}{T_K(N{=}4)} + O\left(\left(\frac{\Delta}{T_K(N{=}4)}\right)^2\right) \qquad (19)$$

The numerical results presented in Fig. 2 reproduce the first term in eq. (19) and agree in the sign of the linear term as well as in the scaling with $(18\hat{g}^2-1)$. However, the numerical pre-factor gleaned from the numerical results is approx. a factor of 13 larger than that in eq. (19).

The scaling behavior for large values of $\Delta/T_K(N{=}4)$ of both $\gamma_0$ and $\chi(T{=}0)$ is governed by eq. (18) so that the Wilson relation [26] holds irrespective of *a*:

$$\frac{4\pi^2}{3}\frac{\chi(T{=}0)}{(g\mu_B)^2\gamma_0} = 2 \quad \text{for } \Delta/T_K(N{=}4)\to\infty \qquad (20)$$

## 5. Apparent non-Fermi-liquid behavior

Fermi-liquid behavior is characterized in the present context by finite values (proportional to the inverse Kondo temperature) of the magnetic susceptibility $\chi$ and the linear specific heat coefficient $\gamma$ for T = 0 with low temperature corrections proportional to $T^2$. The authors of ref. [27] take “the flattening off of the magnetic susceptibility curve below 1 K in contrast with the increasing C/T values…difficult to reconcile with a local Fermi-liquid picture”.

An interpretation of the experimental results of ref. [27] in terms of a local broken SU(4) Anderson model with infinite-U was given in ref. [18] and [21]. In the Kondo limit: T << bandwidth that model is equivalent to the Coqblin Schrieffer model considered here. Consistent quantitative agreement with the results published in Fig. 2 in ref. [21] is obtained with the crystal and magnetic field configuration (c) and the values $\Delta/T_K(N{=}4) = 1.5$ and $T_K(N{=}4) = 7$ K.

In Fig. 3 the specific heat divided by temperature C/T and susceptibility $\chi$ normalized by their respective zero temperature values are shown as functions of temperature T for the three cases of crystal field environments considered and four intermediate crystal field splittings. The small peak at non-zero temperature that is a feature of the SU(4) Coqblin-Schrieffer model is suppressed or enlarged respectively as a function of $\hat{g}$ (or *a*) for small crystal field splittings. On increasing $\Delta/T_K(N{=}4)$ the peak is shifted to higher values of temperature in case (b). In case (a) the peak decreases in height and turns into a shoulder on increasing the crystal field splitting. A similar behavior is found for the specific heat divided by temperature. In case (c) there is no such feature in the susceptibility in the range of crystal field splittings considered.

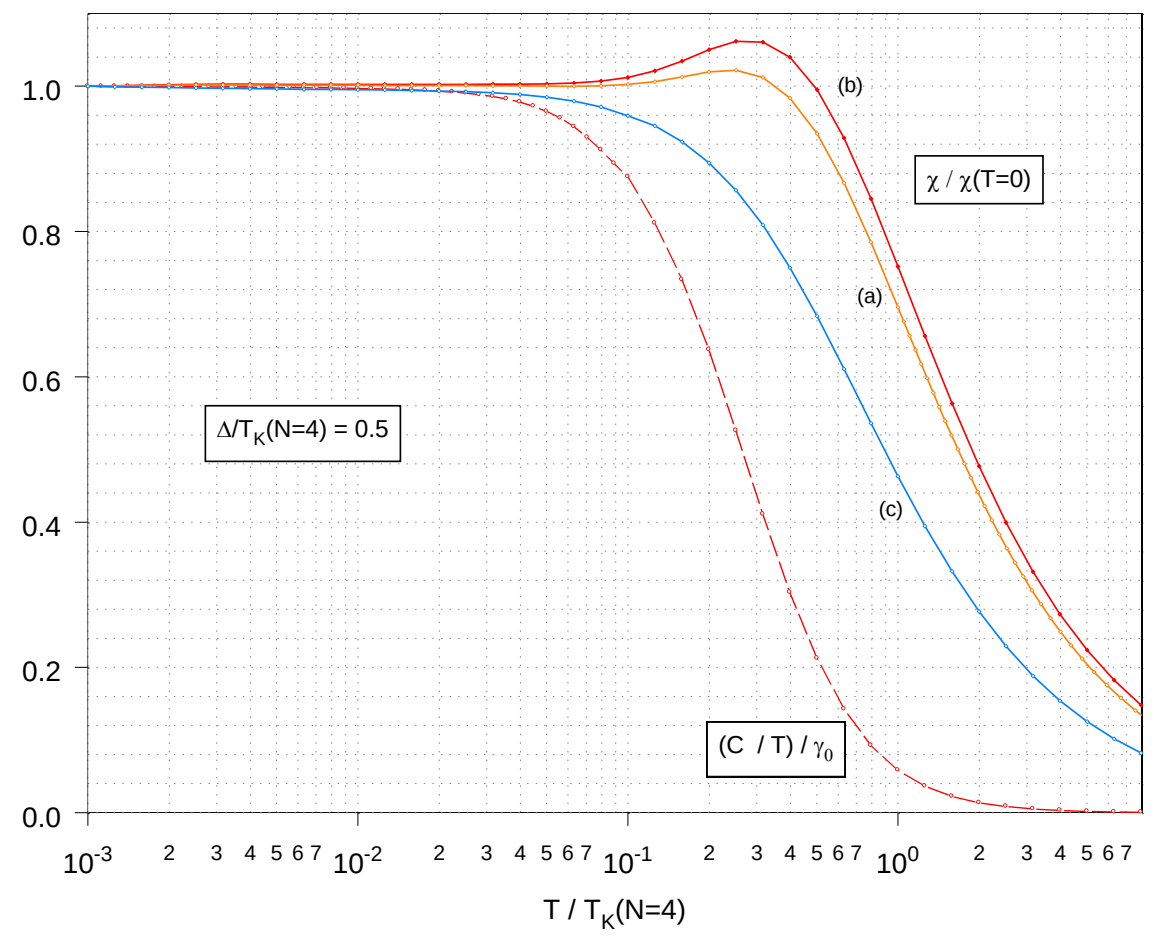

I)

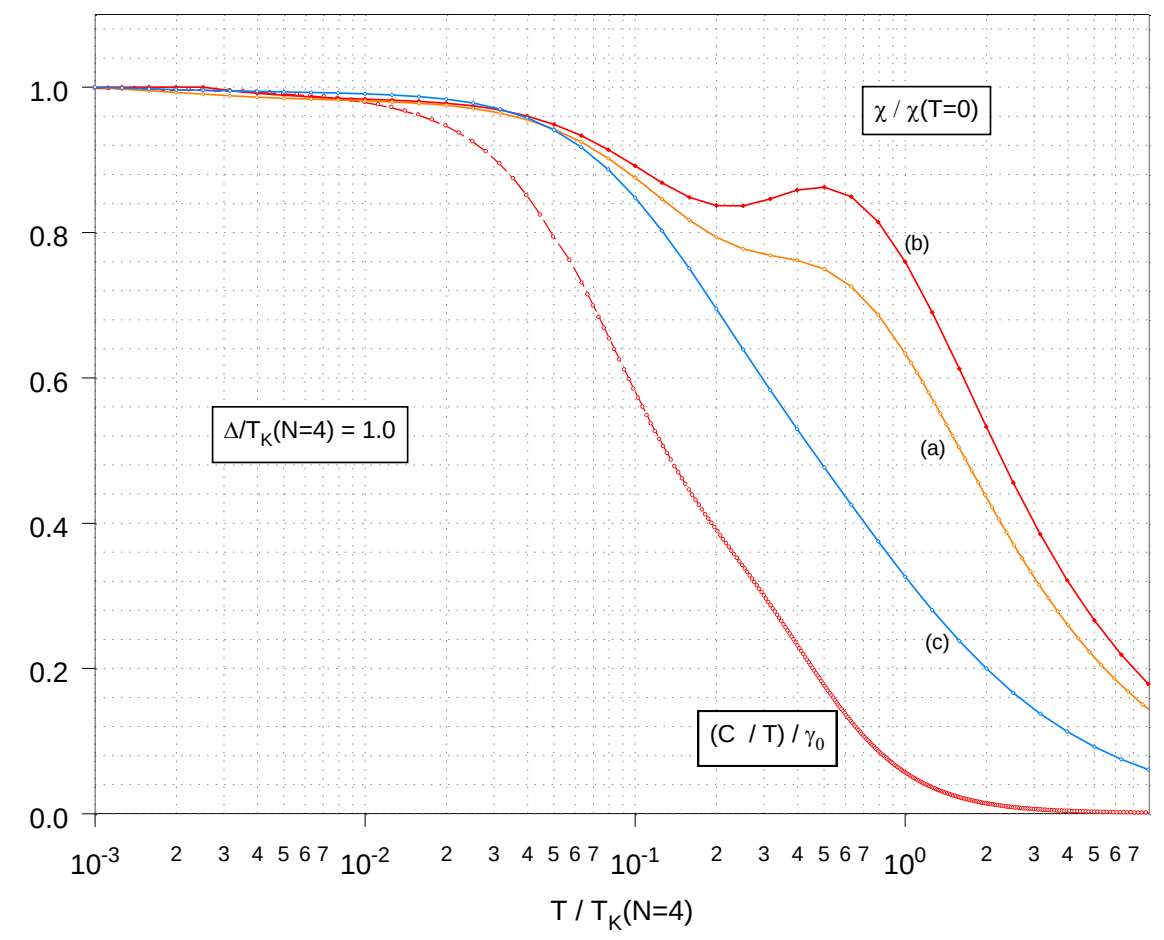

II)

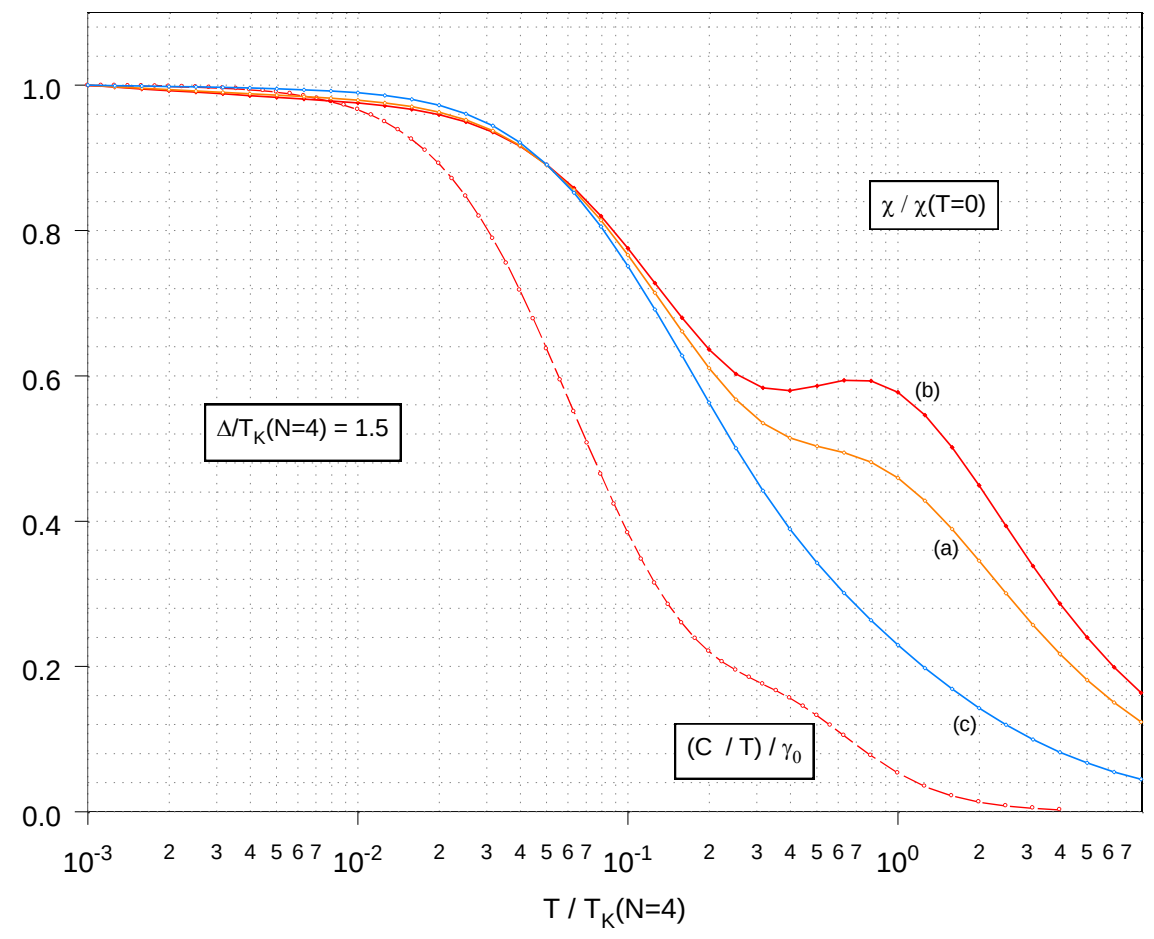

III)

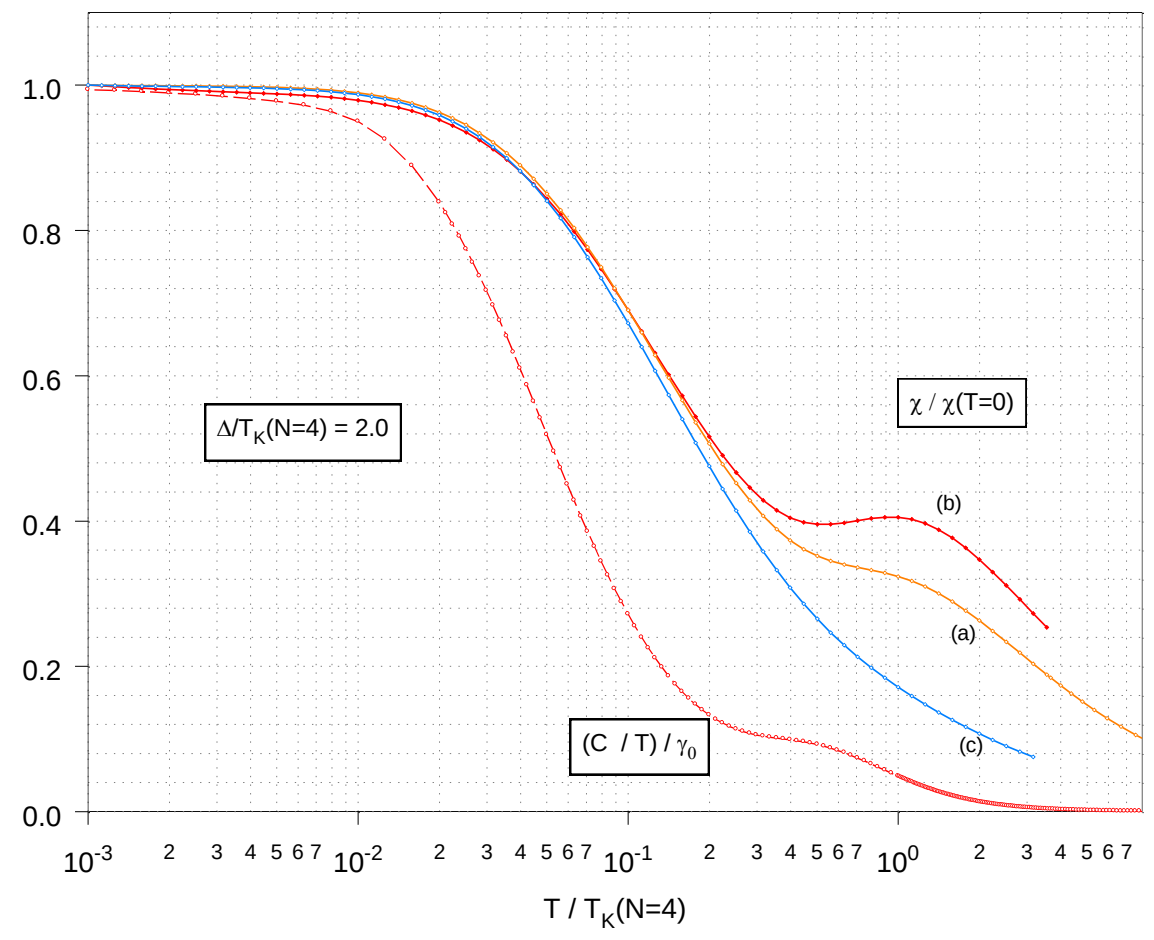

IV)

Fig. 3. (Color online) Specific heat divided by temperature C/T and susceptibility χ normalized by their respective zero temperature values as functions of temperature T for the three cases of crystal field environments considered.

The flattening off of the magnetic susceptibility at higher temperatures than that of the specific heat divided by temperature can also be seen in Fig. 4 where the crossover from medium to large crystal field splittings is shown prototypically for the uniaxial crystal field case (a). The temperature axis is scaled by the effective Kondo temperature $T_K(SU(2))$ [24]. Upon increasing the crystal field splitting the shoulder moves to higher temperatures and decreases in height so that the curve for the SU(2) model is assumed asymptotically.

In Fig. 5 the specific heat divided by temperature C/T and susceptibility χ normalized by their respective zero temperature values are shown for the SU(2) and SU(4) Coqblin-Schrieffer models without crystal fields. The temperature axis is scaled by the respective Kondo temperature $T_K(SU(N))$. It is apparent that the flattening off at different temperatures is a general feature of the SU(2) and SU(4) Coqblin-Schrieffer models. The curves could have been gleaned (in principle) from the original publication of Rajan [28] where the specific heat and susceptibility curves are plotted. However, the characteristic behavior, that becomes obvious in the way it is presented here, has apparently not been recognized before.

Figs. 3 and 4 corroborate the statement from ref. [18]: “The ground state doublet dominates the magnetic response at low temperature and tends to saturate at temperatures higher than the γ-coefficient, consistent with the experiments”. This statement holds also for the crystal and magnetic field configurations (a) and (b). However, as can been seen from Fig. 3, 4, and 5 there is disagreement with the following statement from ref. [18]: “We find this behavior only for CEF splittings $\Delta \approx T^*(\Delta)$ while for much larger or much smaller values $\chi(T)$ and $\gamma(T)$ [i.e. C(T)/T] saturate simultaneously.”

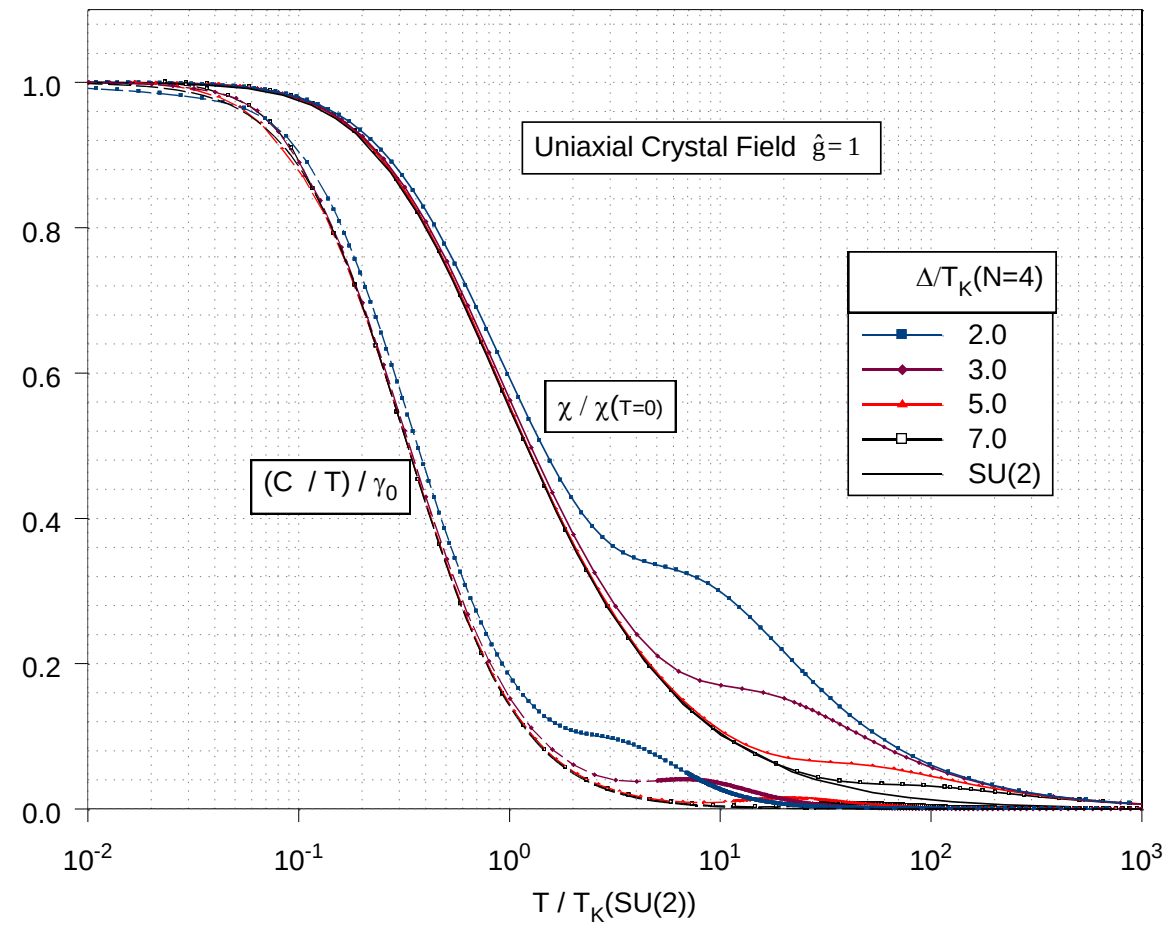

Fig. 4. (Color online) Specific heat divided by temperature C/T and susceptibility χ normalized by their respective zero temperature values as functions of temperature T scaled by the effective Kondo temperature $T_K(SU(2))$ for the uniaxial crystal field case (a), i.e. $\hat{g} = 1$.

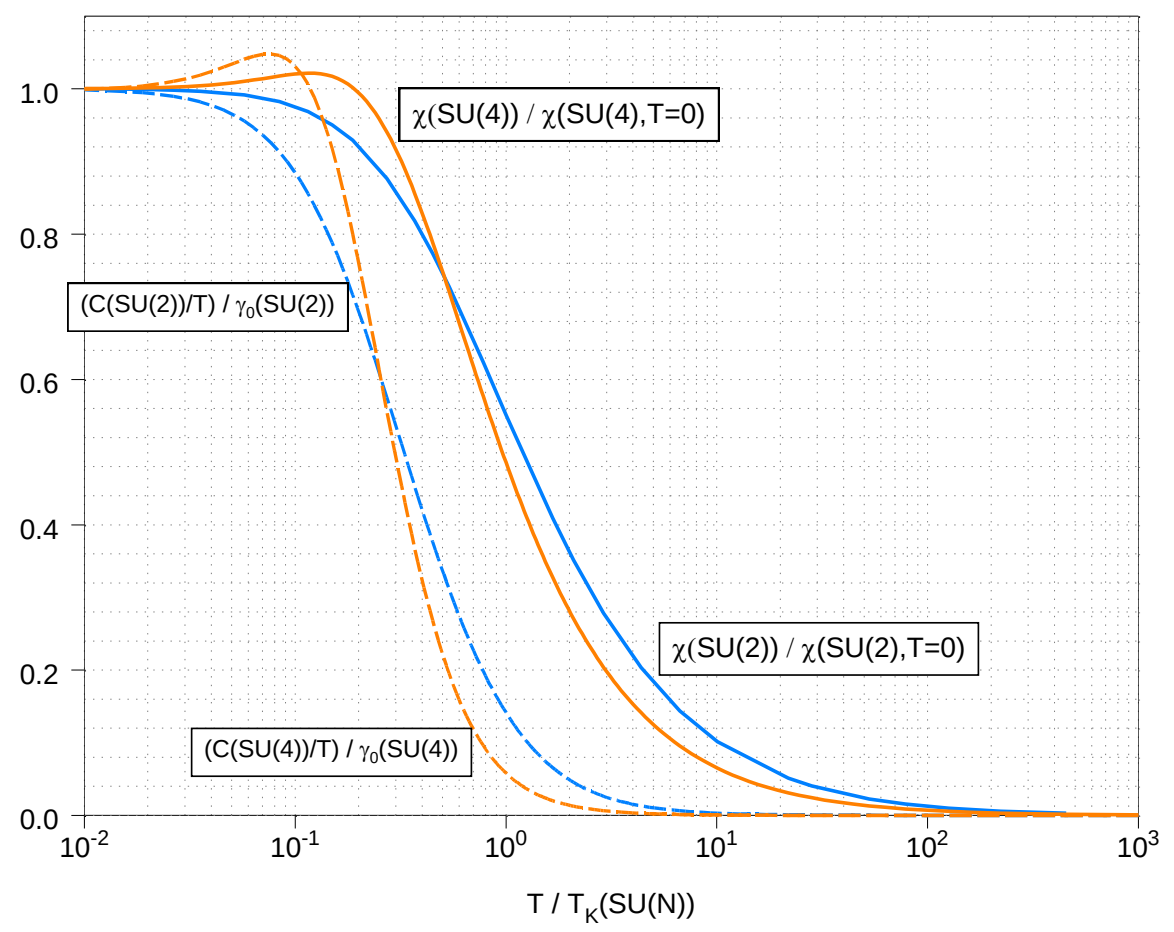

Fig. 5. (Color online) Specific heat divided by temperature C/T and susceptibility χ normalized by their respective zero temperature values as functions of temperature T scaled by the respective Kondo temperature $T_K(SU(N))$ for the SU(N) models without crystal fields for N=2 and N=4.

## 6. Summary and conclusions

The exact Bethe-ansatz solution of the Coqblin-Schrieffer model, a generalization of the Kondo Hamiltonian to include orbital degeneracy N, has been used to provide material for a quantitative analysis of experimental results. The infinite set of coupled, nonlinear integral equations describing the thermodynamics of the model has been solved for the first time to calculate the magnetic susceptibility in the presence of crystal fields for non-zero temperatures.

The numerical solution requires the knowledge of the limiting values $\bar{g}_n^{(r)}$ for the unknown functions $g_n^{(r)}(\lambda)$ of these integral equations that have been known analytically only in the pure magnetic field case. By applying a new result [17] obtained for the N = 6 case with zero magnetic field to the N = 4 case it has been possible to find an analytic formula for $\bar{g}_1^{(r)}$ for zero and infinitesimal magnetic field. While the obtained formulae hold for any crystal field configuration the actual calculation has been done for three prototypical cases. The Bethe-ansatz solution allows for a detailed numerical investigation of the thermodynamic properties with relatively little computational effort.

In this publication the full range of crystal field splittings has been covered. At large splittings the transition to the effective SU(2) (spin ½) behavior characterized by the scaling of specific heat and magnetic susceptibility with the respective SU(2) Kondo temperature is shown. At intermediate crystal fields the different behavior of the three prototypical crystal field configurations considered is investigated.

Case (b) describes a $\Gamma_8$ quartet with tetragonal distortion and magnetic field applied along the z-axis. Among the three cases considered it features the largest Zeeman splitting at a given magnetic field and consequently the largest Curie constant $C \equiv a \cdot \frac{5}{4}(g\mu_B)^2$ with $a$ = 13/9 in this case. By increasing the crystal field splitting the small peak in the magnetic susceptibility at non-zero temperature that is a feature of the SU(4) Coqblin-Schrieffer model is shifted to higher temperatures. It decreases in height but remains a distinct peak.

In case (a) i.e. a spin J = 3/2 ion in a uniaxial crystal field applied along the fourfold axis the Curie constant has its free spin value $\frac{5}{4}(g\mu_B)^2$ (i.e. $a$ = 1). The small SU(4) peak transforms into a shoulder upon increasing the crystal field splitting.

The configuration (c) considered in ref. [18] and [21] has a rather small Curie constant $C \equiv a \cdot \frac{5}{4}(g\mu_B)^2$ with $a$ = 3/10. Here the magnetic susceptibility shows neither peak nor shoulder for the crystal fields considered.

In all cases the flattening off of the magnetic susceptibility occurs at higher temperatures than that of the specific heat divided by temperature. This turns out to be a general feature of the model in contrast to claims that this holds only for a narrow range of crystal field splittings.

*All that glisters is not gold*: Claims of non-Fermi-liquid behavior measured over one decade of temperature may be explained within the well-known Fermi liquid compliant models if the relevant features are incorporated [29].

**Acknowledgment**

I thank Alex Hewson for helpful comments.

**Appendix**

The problem of determining the limiting values in zero magnetic field $\bar{g}_n^{(r)}$ was solved in a previous publication [17] for the N=6 Coqblin-Schrieffer model with the ionic ground state split into three doublets. Letting the splitting between the second doublet and the highest doublet go to infinity one arrives at the solution to (8) for the present N = 4 model for nΔ/T >> 1:

$$\begin{aligned}
&\bar{g}_n^{(1)} = \bar{g}_n^{(3)} = 2\ln[n+1+\alpha] + O(\exp(-(n+1)\Delta/\mathrm{T})),\\
&\bar{g}_n^{(2)} = n\,\Delta/\mathrm{T} - \tfrac{1}{2}\{g_n^{(1)} + g_n^{(3)}\} + f(\Delta/\mathrm{T}) + O(\exp(-(n+1)\Delta/\mathrm{T})),\\
&\alpha = \frac{-2\exp(-\Delta/\mathrm{T})}{1-\exp(-\Delta/\mathrm{T})},\; f(\Delta/\mathrm{T}) = -4\ln[1-\exp(-\Delta/\mathrm{T})], \text{valid for H=0.}
\end{aligned} \tag{21}$$

The numerical procedure to determine the $\bar{g}_n^{(r)}$ follows the same lines as in ref. [17]. To validate the conjectured expression for $\bar{g}_1^{(r)}$ stated in eqs. (15) I have calculated the corresponding expressions for $\bar{g}_n^{(r)}$ up to n = 4 analytically and found agreement with eqs. (21). In order to visualize the behavior of the functions $\bar{g}_n^{(r)}$ approaching their limiting values for large $n$ given by eq. (21) I define α(n) through $\bar{g}_n^{(1)} = 2\ln[n+1+\alpha(\mathrm{n})]$ and plot in the main part of Fig. 6 the numerically obtained α(n) as functions of x for various values of n. The basis of the numerical procedure lies in the fact that for any value of Δ/T there is a reasonably small value of n so that α(n) can be approximated by α.

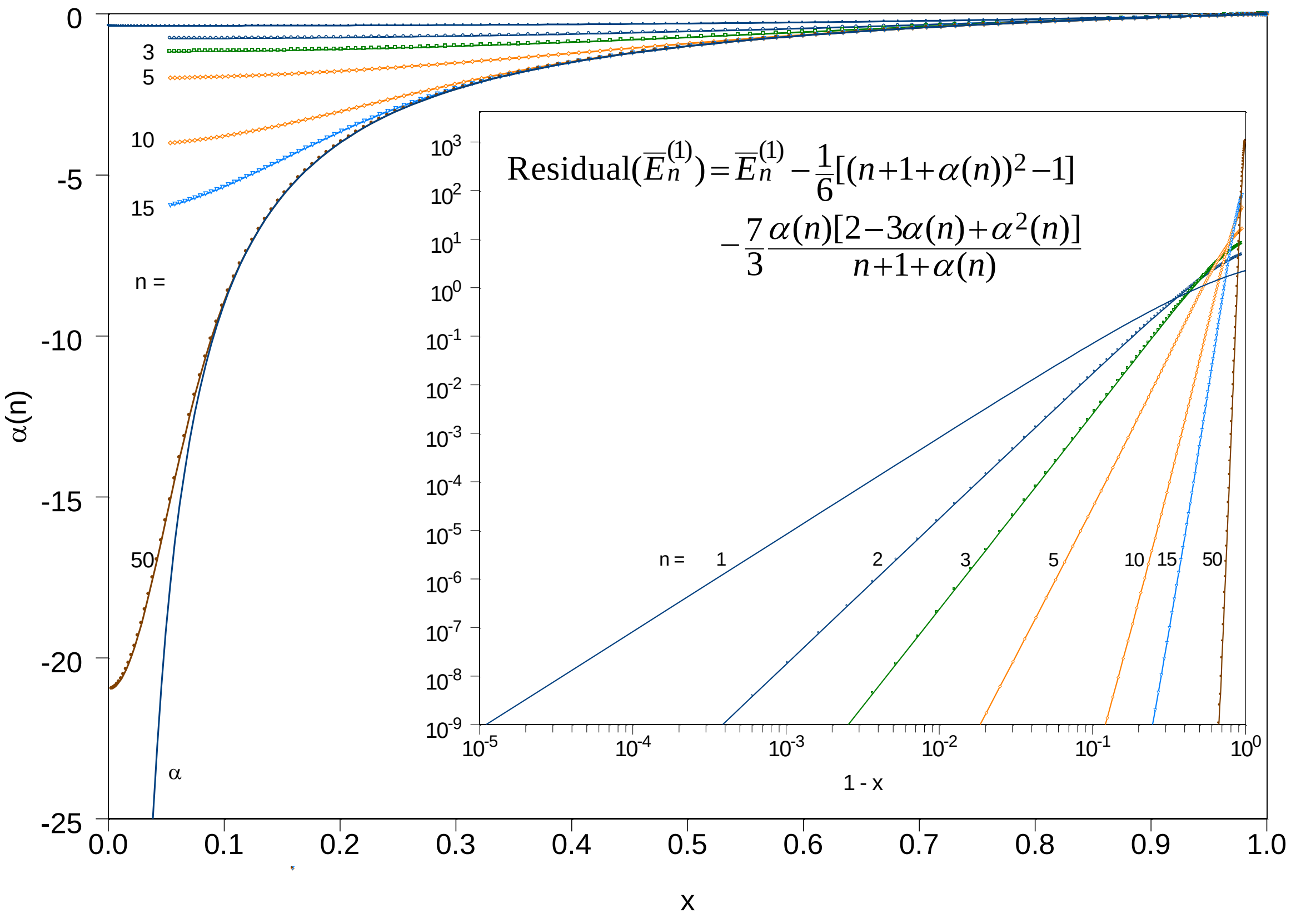

Fig. 6. (Color online) α(n) as functions of x (see text for details)

In a first step, in order to generalize the result (15) to the case H≠0 I made the *ansatz* for large values of nΔ/T:

$$\overline{g}_n^{(t)} = 2\ln\left[\frac{\sinh\{(nA_t/2) + \mathrm{Arsinh}[\sinh(A_t/2)\cdot\{1+\alpha^{(t)}(\Delta/\mathrm{T},\mathrm{H/T})\}]\}}{\sinh\{A_t/2\}}\right] \quad \text{for } t = 1,3, \text{and} \tag{22}$$

$$\overline{g}_n^{(2)} = n\Delta/\mathrm{T} - \tfrac{1}{2}\{g_n^{(1)} + g_n^{(3)}\} + f(\Delta/\mathrm{T},\mathrm{H/T}). \tag{23}$$

With this expressions and by approximating $\alpha^{(t)}(\Delta/\mathrm{T},\mathrm{H/T})$ by $\alpha^{(t)}(\Delta/\mathrm{T},0) \equiv \alpha$ as a second step the limit $\Delta/\mathrm{T}\to\infty$ is reproduced correctly. The limiting values $\overline{g}_n^{(r)}$ were calculated for $H/T_K = 0.001$ from eqs. (8) by employing the same numerical procedure [17] as for H = 0. The function $f(\Delta/\mathrm{T},\mathrm{H/T})$ was approximated by the value obtained from eq. (23) at half the maximum value of n.

In a third step I calculated the limiting values to the ancillary functions $\overline{E}_n^{(r)} = \partial^2\overline{g}_n^{(r)}/\partial^2(\mathrm{H/T})^2\big|_{\mathrm{H}=0}$ by approximating the second derivative by the corresponding difference expression.

From the numerical results the following expression for $\overline{E}_n^{(r)}$ for $n >> -\alpha$ is obtained that may serve as the starting point for the quest for a solution of eqs. (8) for general values of the magnetic field:

$$
\begin{aligned}
\overline{E}_n^{(1)} &= \frac{1}{6}[(n+1+\alpha)^2-1] + \frac{(15a-1)}{6}\frac{\alpha(2-3\alpha+\alpha^2)}{n+1+\alpha} + O(\exp(-(n+1)\Delta/\mathrm{T})), \\
\overline{E}_n^{(2)} &= -\tfrac{1}{2}\{\overline{E}_n^{(1)} + \overline{E}_n^{(3)}\} - 5a\alpha(1-\alpha/2) + O(\exp(-(n+1)\Delta/\mathrm{T})), \qquad (24) \\
\overline{E}_n^{(3)} &= \frac{(3\hat{g})^2}{6}[(n+1+\alpha)^2-1] + \frac{(15a-(3\hat{g})^2)}{6}\frac{\alpha(2-3\alpha+\alpha^2)}{n+1+\alpha} + O(\exp(-(n+1)\Delta/\mathrm{T})).
\end{aligned}
$$

Interestingly enough, the range of convergence of the asymptotic expression (24) can be extended if in these equations α is replaced by α(n). In the insert of Fig. 6 the residual values of $\overline{E}_n^{(1)}$ according to eq. (24) with α replaced by α(n) are displayed as functions of 1-x on a double logarithmic scale for case (a) showing that the residual values are ~ exp(-(n+1)Δ/T) for exp(-Δ/T) ≤ 0.03 n.